# Optical properties of dense zirconium and tantalum diborides for solar thermal absorbers

Elisa Sani*[a], Luca Mercatelli[a], Marco Meucci[a], Andrea Balbo[b], Clara Musa[c], Roberta Licheri[c], Roberto Orrù[c], Giacomo Cao[c]

[a] CNR-INO National Institute of Optics, Largo E. Fermi, 6, I-50125 Firenze, Italy
[b] Corrosion and Metallurgy Study Centre "Aldo Daccò", Engineering Department, University of Ferrara, G. Saragat 1, Ferrara 44122, Italy
[c] Dipartimento di Ingegneria Meccanica, Chimica e dei Materiali, Unità di Ricerca del Consorzio Interuniversitario Nazionale per la Scienza e Tecnologia dei Materiali (INSTM), Università degli Studi di Cagliari, Piazza D'Armi, 09123 Cagliari, Italy
* Corresponding author, email: elisa.sani@ino.it

**ABSTRACT**

Ultra-high temperature ceramics (UHTCs) are interesting materials for a large variety of applications under extreme conditions. This paper reports on the production and extensive characterization of highly dense, pure zirconium and tantalum diborides, with particular interest to their potential utilization in the thermal solar energy field. Monolithic bulk samples are produced by Spark Plasma Sintering starting from elemental reactants or using metal diboride powders previously synthesized by Self-propagating High-temperature Synthesis (SHS). Microstructural and optical properties of products obtained by the two processing methods have been comparatively evaluated. We found that pure diborides show a good spectral selectivity, which is an appealing characteristic for solar absorber applications. No, or very small, differences in the optical properties have been evidenced when the two investigated processes adopted for the fabrication of dense $TaB_2$ and $ZrB_2$, respectively, are compared.

## 1. Introduction

The so-called "energy problem" is one of the major challenges to address in the next future. Solutions will compulsorily include renewable energies, and in particular solar energy exploitation, because energy must be supplied to a growing world population in a safe, environmental-friendly and sustainable way. In this regard, solar thermal technology (1) shows efficiencies intrinsically higher than photovoltaic one. The scheme of concentrating the solar power to a central tower (2) is considered one of the most promising, as it additionally can exploit mature technologies of conventional fossil fuels plants. However, it is important to increase the operating temperature of the plants to improve the efficiency of their thermal cycles. In this context, the solar radiation receiver is a critical element. In particular, several requirements have to be satisfied (3-5), like a high solar absorbance, to efficiently absorb sunlight, a low thermal emittance, to limit thermal re-radiation losses, and good thermal properties, to properly transfer the thermal energy to the exchange medium. Thus, the main challenge for increasing the operating temperature of thermal plants is the development of novel receiver materials which are stable at very high temperatures and able to exhibit all the favorable optical and thermal properties mentioned above.

To this aim, particular attention has been recently focused on the group of materials known as Ultra-High-Temperature-Ceramics (UHTCs), which includes carbides, borides and nitrides of early transition metals and are characterized by very high melting temperatures (above 3200K). UHTCs are known to be ideal materials for thermal protection systems, especially those requiring chemical and structural stability at extremely high operating temperatures thanks to their solid state stability, good thermochemical and thermomechanical properties, high hardness, high electrical and thermal conductivities (6-8). In this regard, some relevant structural, thermodynamic, physical, and mechanical properties of two of the most representative UHTCs, namely $ZrB_2$ and $TaB_2$, are summarized in Table 1. Historically, UHTCs are mainly employed in the aerospace industry for hypersonic vehicles, rocket motor nozzles or atmospheric entry probes capable of the most extreme entry conditions (6-8). In addition to this application, in the last few years they have been also proposed as possible candidates for novel solar absorbers able to operate at very high temperature, because of their favorable optical and radiative properties like intrinsic spectral selectivity and low thermal emittance (9-11). In fact, it is well established that spectral selectivity is a key parameter for increasing the efficiency of solar thermal systems (12). In addition to that, it should be also mentioned the importance of a high solar absorbance, which represents, on the other hand, a possible criticism for UHTCs, as this property is generally lower, for instance, than that of silicon carbide (SiC). However, it has been recently demonstrated, for the case of hafnium carbide (11), that surface texturing can selectively increase solar absorbance without detrimentally affecting thermal emittance.

As for borides, optical and structural properties of several materials belonging to this family, including additive containing $ZrB_2$ and $TaB_2$, have been recently investigated and appealing characteristics for solar absorber applications have been found (13-16). Despite of their attractive properties, the refractory character of Zr and Ta borides makes powder consolidation very difficult, particularly when taking advantage of conventional pressure assisted sintering methods, where holding temperatures even above 2300K and prolonged times (hours) are typically needed (6)(17). Nevertheless, materials with residual porosity and coarse microstructure are often obtained under such processing conditions. Therefore, various approaches have been proposed to overcome this problem. In this regard, the use of innovative densification techniques, like the Spark Plasma Sintering (SPS) or other electric current assisted consolidation methods, was demonstrated particularly advisable for processing difficult-to-sinter materials (18). The same technology was also successfully adopted for reactive sintering purpose, through the so-called Reactive SPS (RSPS)(19-22). Furthermore, the SPS conditions can be further mitigated if starting from powders with high sintering ability. To this aim, it was found that the use of powders prepared by Self-propagating High-temperature Synthesis (SHS) leads to higher density samples, with respect to materials with the same nominal composition obtained using alternative preparation routes (23, 24).

In this work, the optical properties of dense and monophasic $ZrB_2$ and $TaB_2$ are investigated for the first time, to the best of our knowledge. Both borides have been produced using the two different RSPS and SHS/SPS techniques. Microstructural and optical characteristics of the resulting products have been compared.

| **Property** | **$ZrB_2$** | **$TaB_2$** |
|---|---|---|
| Crystal system space group prototype structure<br>a (Å)<br>c (Å) | Hexagonal<br>P6/mmm $AlB_2$<br>3.17<br>3.53<br>(6) | Hexagonal<br>P6/mmm $AlB_2$<br>3.0880<br>3.2410<br>(25) |
| Density (g/cm$^3$) | 6.119<br>(6) | 12.568<br>(25) |
| Enthalpy of formation, at 25°C (kJ/mol) | -322.6<br>(6) | -209.200<br>(26) |
| Free Energy of Formation, at 25°C (kJ) | -318.2<br>(26) | -206.7<br>(26) |
| Melting Temperature (°C) | 3245<br>(6) | 3037<br>(27) |
| Coefficient of Thermal Expansion (K$^{-1}$) | $5.9 \times 10^{-6}$<br>(6) | $8.2 \times 10^{-6}$<br>(28) |
| Heat Capacity, at 25°C (J·(mol·K)$^{-1}$) | 48.2<br>(6) | 33.92<br>(29) |
| Electrical conductivity, at 25°C (S/m) | $1.0 \times 10^7$<br>(6) | $3.03 \times 10^6$<br>(30) |
| Thermal conductivity (W·(m·k)$^{-1}$) | 60<br>(6) | 16.0<br>(28) |
| Young's Modulus (GPa) | 489<br>(6) | 248.2<br>(29) |
| Hardness (GPa) | 23<br>(6) | 19.6<br>(31) |

**Table 1.** Properties of $ZrB_2$ and $TaB_2$.

## 2. Experimental procedure

A Spark Plasma Sintering (SPS) equipment (515 model, Sumitomo Coal Mining Co Ltd, Japan) was used under mild vacuum conditions (20 Pa) to obtain $TaB_2$ and $ZrB_2$ dense samples by reactive (RSPS) or non-reactive sintering. In the latter case, transition metal diborides were preliminarily synthesized by SHS and the obtained powders consolidated by SPS. Such two steps processing route will be hereafter indicated as SHS/SPS. Initial mixtures consist of B (Sigma-Aldrich, amorphous, particle size < 1 μm, ≥ 95% purity), Zr (Alfa Aesar, particle size < 44 μm, 98.5% purity) or Ta (Alfa-Aesar, particle size < 44 μm, 99.6% purity) combined according to the reaction Me + (2+x) B → $MeB_2$, where Me = Zr or Ta. The use of a slight excess of B (x = 0.1) with respect to the stoichiometric value was aimed to remove some oxide impurities originally present in the raw materials, particularly in Zr powders. SHS reactions, performed under Ar atmosphere, were locally activated using an electrically heated tungsten coil. Further details on the SHS procedure can be found elsewhere (32). The obtained porous samples were ball milled for 20 min to provide powders with average particle sizes of approximately 6.7 and 1.5 μm for $ZrB_2$ and $TaB_2$, respectively.

Dense samples with 14.7 mm diameter and about 3 mm thickness were produced by SPS starting from proper amounts of ZrB$_2$ and TaB$_2$ powders (SHS/SPS) or elemental reactants (RSPS). During the consolidation process, the applied current was increased from zero to $\bar{I}$ = 1300 A in 10 min. Then, the $\bar{I}$ value was maintained constant for additional 20 minutes. The applied mechanical pressure was generally set in the range 50-60 MPa, except for the case of the ZrB$_2$ system processed by RSPS, where relatively lower loads (20 MPa) were used. This was to avoid some drawbacks (rapid gas development, product inhomogeneity, etc.) and safety problems (die/plunger breakage, etc.) encountered when the strongly exothermic reaction for the ZrB$_2$ formation from its elements was allowed to evolve in the combustion regime in presence of pressure levels above 20 MPa (22). Additional information related to SPS experiments are reported elsewhere (20)(22).

The crystalline phases in SHS and RSPS products were identified by X-ray diffraction using a Philips PW 1830 X-rays diffractometer with a Ni filtered Cu K$_\alpha$ radiation ($\lambda$=1.5405 Å). Dense samples were ground and polished with diamond pastes down to 0.25 μm.

The surface microstructure of bulk materials was then examined by scanning electron microscopy using a ZEISS EVO LS 15 apparatus equipped with a LaB$_6$ filament as electron source. Relative densities of sintered products were evaluated using the Archimedes' method, by considering the theoretical values for ZrB$_2$ and TaB$_2$ equal to 6.1 (33) and 12.6 g/cm$^3$ (29), respectively.

The topological characterization of the surfaces was carried out with a non-contact 3D profilometer (Taylor-Hobson CCI MP) on two areas of 0.08 x 1 cm$^2$ at the center of each sample and the topography data were analysed using a commercial software (Talymap 6.2).The evaluation of 2D texture parameters (Ra, Rt) was performed on 4 different profiles (2 for each area) extracted from the 3D data, while the gaussian filter ($\lambda$c) for the separation of the roughness and waviness components was set according to the ISO 4288:2000. The 2D parameters were calculated as average of estimated values by considering all sampling lengths over each profile.

Hemispherical reflectance spectra were acquired with two instruments: a double-beam spectrophotometer (Lambda 900 by Perkin Elmer) provided with a Spectralon©-coated integration sphere for the 250-2500 nm wavelength region and a Fourier Transform spectrophotometer (FT-IR "Excalibur" by Bio-Rad) equipped with a gold-coated integrating sphere and a liquid nitrogen-cooled detector for the range 2500-16500 nm.

## 3. Results
### 3.1 Surface microstructure

XRD patterns shown in Figure 1 indicate that, during both SHS and RSPS processes, the initial elemental reactants are fully converted to the desired boride phases. The density values of the diboride products obtained by SPS are reported in **Table 2**. As far as ZrB$_2$ is concerned, relatively higher density samples are obtained when using SHS powders with respect to specimens processed by RSPS. This feature confirms the high sintering ability of ZrB$_2$ powders produced by SHS (23). However, as mentioned in the previous section, it should be recalled that a relatively lower mechanical pressure was adopted during reactive sintering with respect to that applied when processing SHS powders, i.e. 20 and 50 MPa, respectively.

On the other hand, when considering the TaB$_2$ system, the RSPS process was able to provide slightly higher density products, as compared to the SHS/SPS route. The occurrence of the rapid combustion synthesis reaction during reactive SPS was apparently beneficial in the latter case (20).

The SEM micrographs reported in Figure 2 relative to the polished surfaces of dense samples obtained by RSPS and SHS/SPS are in agreement with density measurements. Indeed, although the microstructures show that a good level of consolidation is reached, the presence of a residual amount of closed porosity is also evidenced. In particular, $ZrB_2$ samples obtained by RSPS display a relatively coarser microstructure compared with specimens obtained using the alternative route, with grains size up to 50 and 20-30 μm, respectively. In addition, large pores (up to 6-8 μm sized) are produced during the reactive sintering process. This feature could be likely associated to the gas development during the *in situ* occurrence of the synthesis reaction, caused by the presence of some oxide impurities in the initial reactants. Indeed, such gases can be released during the evolution of the SHS reaction in free-standing pellets, whereas they are not allowed to escape from the graphite container during RSPS, so that they remain entrapped within the processing samples. Figure 2 shows that relatively finer and similar microstructures with grains at most 10 μm sized are obtained for the case of the $TaB_2$ system processed by the two SPS methods. In addition, the sintered material exhibits relatively smaller pores distributed throughout the material, albeit the total porosity is rather high, as revealed by density measurements. Also in this case, relatively larger pores are generated when taking advantage of the RSPS route. The coarser ($ZrB_2$) and finer ($TaB_2$) microstructures of massive products obtained using the non reactive sintering routes can be readily associated to the corresponding particle sizes of the starting SHS powders, i.e. 6.7 and 1.5 μm, respectively.

The considerations made above hold also true when examining the bulk of the sintered materials. Indeed, from the fracture surfaces of Zr and Ta diboride specimens shown in Figure 3, it is confirmed that the products obtained by RSPS display larger pores with respect to those one produced by SHS-SPS. Furthermore, pore sizes of ceramic bulk are even larger, particularly for the case of $ZrB_2$ prepared by reactive sintering, than those ones observed on the sample surface (Figure 2). This is because gases developed during the synthesis reaction near to the material surface can leave the sample more easily, as compared to those ones formed in the center region. Even though sufficiently high density values are achieved in the present work for both systems and fabrication methods, some useful considerations can be made, on the basis of the results described above, to further improve the final consolidation level in future investigations. Specifically, for the case of the RSPS process, it is expected that the use of higher purity elemental powders will reduce the amount of gases liberated during the synthesis reaction and, in turn, the resulting sample porosity. In addition, the fabrication of SHS powders with grain size down to the nanometric level, which is made possible either using appropriate additives during the synthesis process or through a high-energy ball milling treatment of the obtained product, will also likely provide an additional contribution to improve their densification.

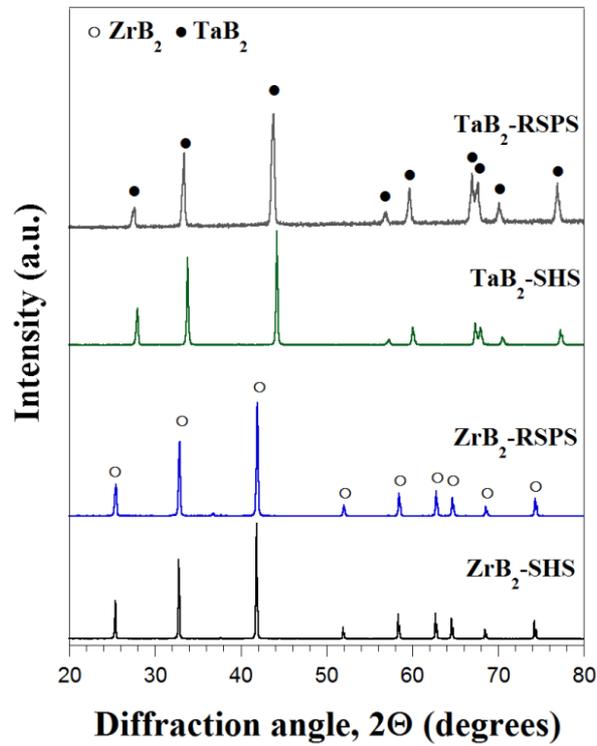

**Figure 1**. XRD patterns of Zr- and Ta-diboride products obtained after SHS and RSPS.

| System | Method | Relative density (%) | Grain size (µm) | Pore size (µm) |
|---|---|---|---|---|
| ZrB$_2$ | RSPS | 96.0±0.4 | 50 | 6-8 |
|  | SHS/SPS | 98.3±0.8 | 20-30 | 3-4 |
| TaB$_2$ | RSPS | 95.4±0.7 | 10 | 1.5-3.5 |
|  | SHS/SPS | 94.0±0.4 | 10 | 0.5-3.0 |

**Table 2**. Characteristics of the obtained Zr- and Ta-diborides.

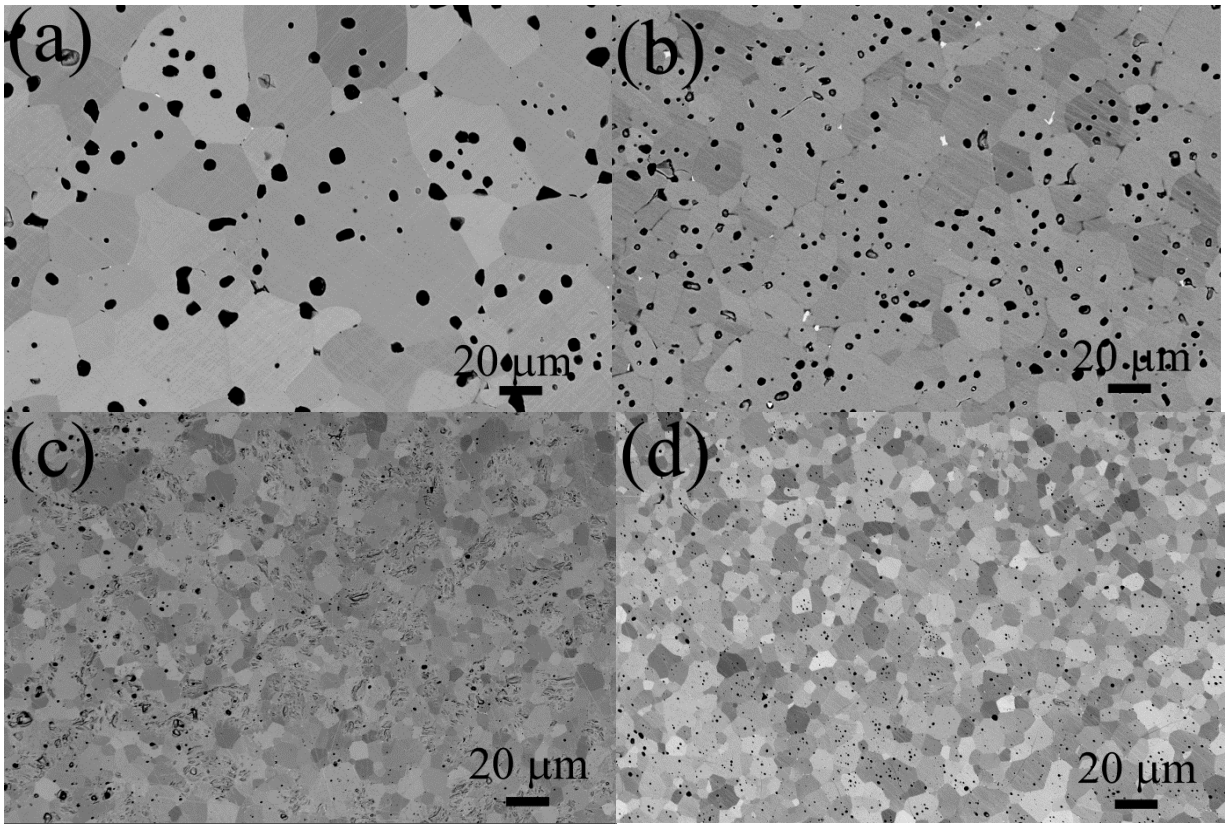

**Figure 2.** SEM micrographs of polished surfaces of SPS products exposed to optical measurements: (a) **ZrB₂** RSPS and (b) **ZrB₂** SHS/SPS, (c) **TaB₂** RSPS and (d) **TaB₂** SHS/SPS.

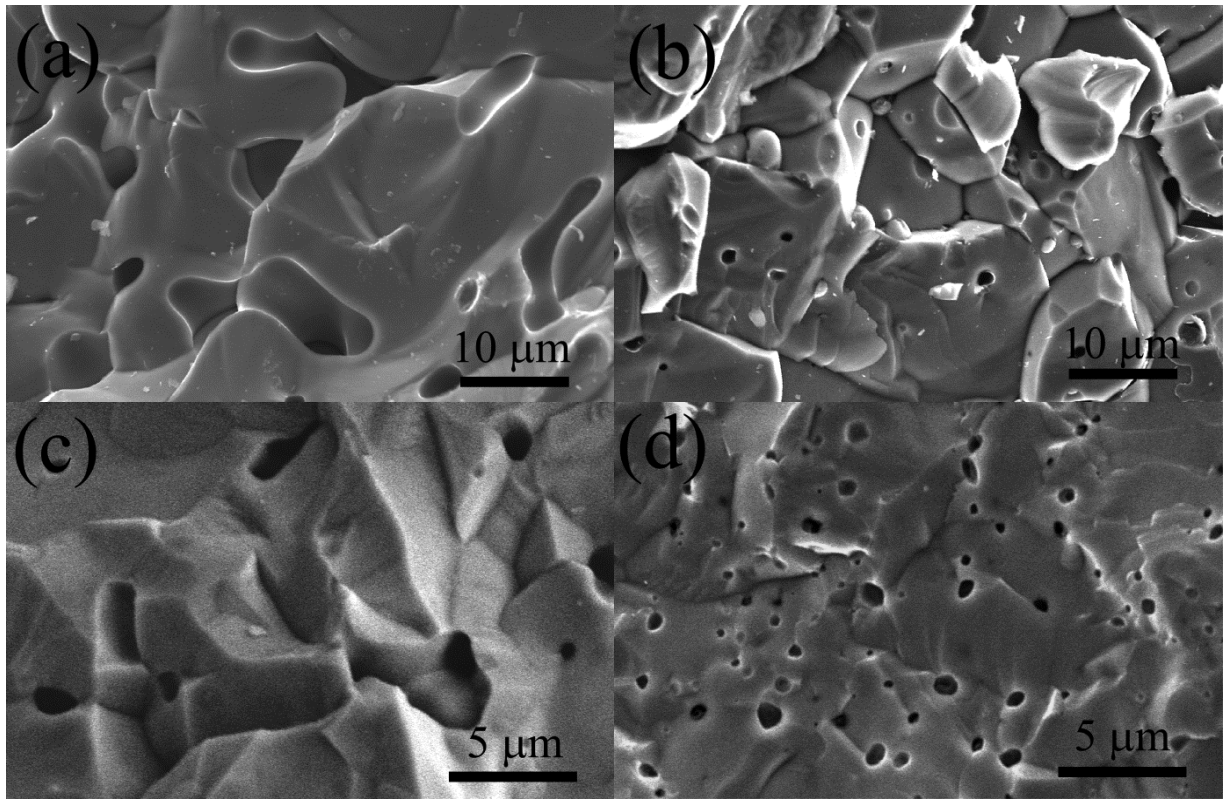

**Figure 3.** SEM micrographs of fracture surfaces showing bulk microstructures of sintered products: (a) **$ZrB_2$** RSPS and (b) **$ZrB_2$** SHS/SPS, (c) **$TaB_2$** RSPS and (d) **$TaB_2$** SHS/SPS.

*3.2    Roughness*

Measured roughness of sintered products is plotted in Figure 4. The $ZrB_2$ sample obtained by SHS/SPS is the smoothest (Ra=58±9 nm, Rt=2400±600 nm), probably owing to the higher density and smaller grain size and pores. On the other hand, $TaB_2$ pellets produced by the two different processing routes show similar roughness characteristics, in agreement with the corresponding similar densities and grain sizes (Ra=100±20 nm, Rt=3400±1700 nm for SHS/SPS sample and Ra=94±6 nm, Rt=2250±300 nm for RSPS sample).

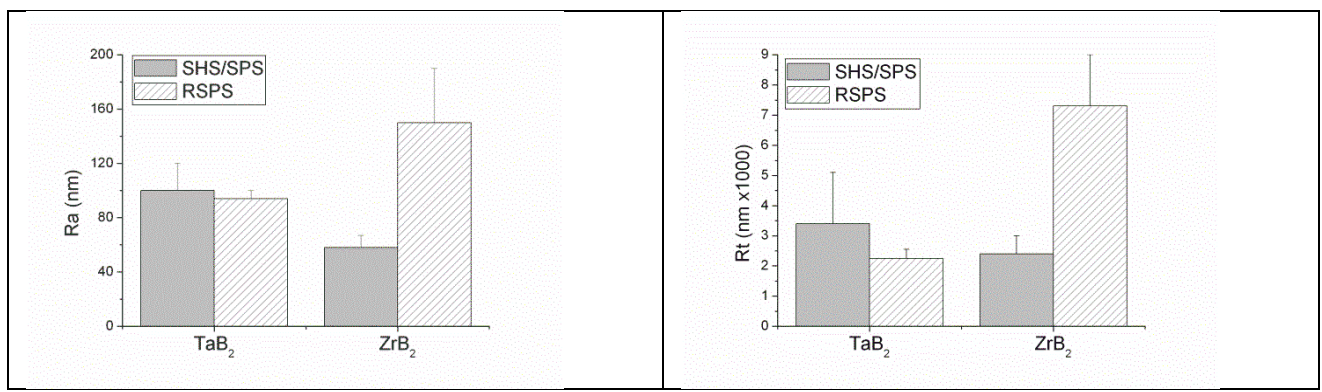

**Figure 4.** Roughness values of dense metal-diboride samples subjected to optical measurements. Ra: average roughness; Rt: maximum distance between peak and valley.

*3.3    Optical properties*

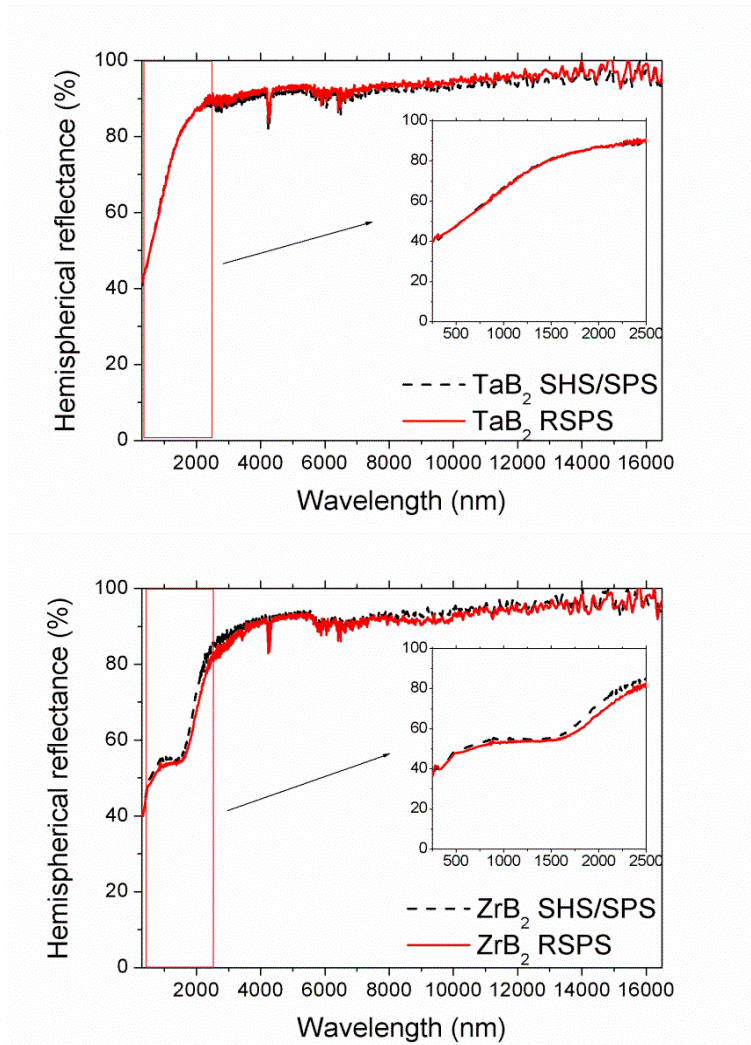

**Figure 5.** Hemispherical reflectance spectra of TaB$_2$ and ZrB$_2$ samples. The inset shows in detail the 250-2500 nm region.

Figure 5 shows the room-temperature hemispherical reflectance spectra. No significant differences have been found in both TaB$_2$ and ZrB$_2$ systems when comparing the utilized production methods. Therefore, the small discrepancies in surface roughness values and pore size, particularly when considering the ZrB$_2$ system, appear to entail a minor effect on optical spectra. As for the influence of roughness, this can be mainly ascribed to differences in the Rt values (which, in turn, are connected to the pore depth) as, in any case, typical Ra ranges for our samples are much lower than the investigated optical wavelengths.

To evaluate spectral differences among the two considered borides, Figure 6 compares the spectra of TaB$_2$ and ZrB$_2$ for fixed SHS/SPS processing technique. Both borides show step-like reflection curves characterized by a quick reflectance increase for wavelengths up to about 3000 nm and nearly constant high reflectance plateau at longer wavelengths. The TaB$_2$ and ZrB$_2$ spectra can be nearly superimposed each other for wavelengths below 550 nm and above 2600 nm. In the intermediate spectral region 550-2600 nm the reflectance of TaB$_2$ is higher, being its curve monotonically increasing, than that of ZrB$_2$, which is characterized by a nearly constant reflectance in the range 850-1400 nm and a monotonic increase from 1400 to 2600 nm.

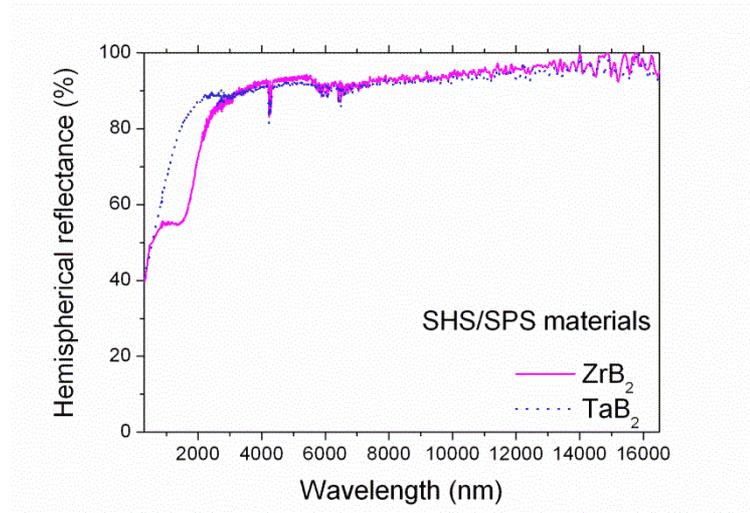

**Figure 6.** ZrB$_2$ and TaB$_2$ spectra (SHS/SPS method).

To evaluate the impact of spectral features on solar absorber characteristics, we calculate the total directional solar absorbance

$$\alpha'_S(T) = \frac{\int_{\lambda_{min}}^{\lambda_{max}} (1-\rho'^{\cap}(\lambda,T)) \cdot S(\lambda) d\lambda}{\int_{\lambda_{min}}^{\lambda_{max}} S(\lambda) d\lambda} \qquad (1)$$

and the total directional emittance

$$\varepsilon'_{\lambda 1,\lambda 2}(T) = \frac{\int_{\lambda 1}^{\lambda 2} (1-\rho'^{\cap}(\lambda,T)) \cdot B(\lambda,T) d\lambda}{\int_{\lambda 1}^{\lambda 2} B(\lambda,T) d\lambda} \qquad (2)$$

being S(λ) the sunlight spectrum (34), $\lambda_{min}$ = 300 nm, $\lambda_{max}$ = 3000 nm, B(λ,T) the blackbody spectrum evaluated at the temperature T, $\lambda_1$ = 300 nm, $\lambda_2$ = 15000 nm. Under the remarks made in (35), the ρ'$^{\cap}$(λ,T) function can be approximated with the spectral hemispherical reflectance measured at room temperature. In this regard, we notice that for assessing the actual material performances under operative conditions, the knowledge of optical properties at high temperatures is required. However, due to the electric current limitation of 1500 A, the SPS 515 apparatus available in the present investigation cannot be used to produce the larger specimens needed for the direct high temperature optical characterization (9) (36, 37). To this aim, a larger SPS equipment able to provide significantly higher electric current intensities has to be used. Nonetheless, this activity will represent the further development of this work, which can be then considered the preliminary step of the complete study.

Moreover, it should be noted that direct high temperature emittance measurements recently carried out on differently produced TaB$_2$ (relative density up to about 90%) and HfC samples (37), have shown that, although the obtained room temperature measurements underestimate the observed value of spectral emittance at high temperature, the estimation of emittance from room temperature reflectance data provides reliable and consistent results, if the comparison among an homogeneous group of samples is concerned. Thus, the hierarchy exhibited by samples at room temperature is also maintained at high thermal levels.

Thus, when, for instance, an operating temperature T=1400 K is considered for our samples, we obtain α = 0.47 for $ZrB_2$ and 0.40 for $TaB_2$ and ε = 0.18 and 0.12 for $ZrB_2$ and $TaB_2$, respectively. In other words, the low-reflectance shoulder evidenced in the $ZrB_2$ spectrum makes such ceramic a better solar absorber with respect to the $TaB_2$ sample. On the other hand, the zirconium diboride product results to be a more emissive material. Finally, if the α/ε ratio at 1400 K (2.6 for $ZrB_2$ and 3.3 for $TaB_2$) is considered, the Ta-based specimen shows a relatively more interesting performance than the $ZrB_2$ one. However, it should be noted that when the required operating temperature is lower, blackbody spectrum is shifted at longer wavelengths, so that the detrimental effect of $ZrB_2$ shoulder on the emittance is correspondingly decreased. For instance, α/ε = 3.9 for $ZrB_2$ and 4.0 for $TaB_2$, respectively, when T = 1000 K. Thus, this preliminary investigation shows that transition metal borides may be interesting as base matrices for solar absorber applications. However, the direct exploitation of these materials will be possible only after a careful optimization, in particular aimed to increase their solar absorbance, and if a proper choice of the plant architecture/operating temperature is adopted.

## 4. Conclusions

Additive-free $ZrB_2$ and $TaB_2$ dense ceramics prepared with the SHS/SPS and RSPS techniques are investigated in this work. Specifically, the obtained samples have been characterized as for surface microstructure, roughness and optical properties, to evaluate their potential utilization for solar absorber applications. When the two processing routes for the preparation of bulk $TaB_2$ are compared, it is found that the measured optical properties of the end products are practically identical. This feature is in agreement with the similar resulting microstructures. When the optical behavior of $ZrB_2$ specimens is examined, small differences in the reflectance values have been detected. This outcome can be mainly ascribed to some dissimilarities in pore size displayed by the two kind of $ZrB_2$ samples, which, in turn, affect the corresponding Rt value of roughness. Nevertheless, both borides show step-like reflectance spectra, characterized by a low reflectance in the visible-near infrared and high reflectance values in the mid-infrared, arising in a high solar absorbance and a low thermal emittance at high temperatures. Spectral differences between $ZrB_2$ and $TaB_2$ are limited to the wavelength range between 550 and 2600 nm and produce an absorbance-to-emittance ratio (α/ε) slightly higher for $TaB_2$ in comparison with that obtained for $ZrB_2$.

Based on the result obtained in the present work, it is possible to conclude that both the RSPS and SHS/SPS routes are capable to lead to massive and monolithic $ZrB_2$ and $TaB_2$ products which exhibit only minor differences in the optical properties. However, some consideration on the practical use of these two processing methods should be made. Indeed, although the RSPS approach has the advantage to allow the synthesis and simultaneous consolidation of the material in one single step, the highly exothermic character of the reactions for the formation of Ta or Zr diborides from the corresponding elemental powders might cause some problems. Indeed, caution should be taken to avoid, or properly control, the occurrence of these reactions under the combustion regime in the confined environment represented by the graphite mould. This certainly represents a crucial aspect to be considered in view of a possible process scale-up for the fabrication of transition metal diborides by reactive SPS.

On the other hand, the separation of the synthesis (SHS) and consolidation (SPS) stages in the alternative route permits to overcome the previously mentioned drawback. Thus, at least for the case of the two systems taken into account in this work, the SHS/SPS approach might be considered more promising for practical exploitation.

In conclusion, some final considerations are needed relatively to the direct utilization of monophasic UHTCs at high temperature levels under oxidative environment. Indeed, it is well known that tantalum and zirconium diborides start to oxidize above approximately 600 °C to produce a rapid material degradation, unless suitable additives are introduced (38). For instance, oxidative tests conducted on $ZrB_2$ coatings developed for possible photothermal solar applications indicate that the maximum operating temperature for air exposure was about 600°C (39). Analogous considerations can be made for $TaB_2$ (40). Thus, no oxidative problems are expected to occur below such temperature limits for the use of monolithic $ZrB_2$ and $TaB_2$ as solar absorber. On the other hand, appropriate amounts of suitable, typically silicon-containing additives (SiC, $MoSi_2$, etc.), have to be added to improve their oxidation resistance and make their utilization possible under more severe thermal conditions.

**Acknowledgements**
This activity has been carried out in the framework of the FIRB2012-SUPERSOLAR (Programma "Futuro in Ricerca", prot. RBFR12TIT1) project funded by the Italian Ministry of Education, University and Research. E. S. gratefully acknowledges the Italian bank foundation "Fondazione Ente Cassa di Risparmio di Firenze" for providing the grant for M.M. within the framework of the "SOLE" project (pratica n. 2013.0726). Authors thank Dr. L. Silvestroni and Dr. D. Sciti (ISTEC-CNR) for useful discussions about sample preparation and microstructural analysis. Thanks are also due to Mr. Mauro Pucci and Mr. Massimo D'Uva for technical assistance.